\documentclass[10pt, conference, compsocconf]{IEEEtran}
\usepackage{multicol}
\usepackage{color}
\usepackage{xspace}
\usepackage{pdfwidgets}
\usepackage{enumerate}
\usepackage{etoolbox}
\usepackage{balance}
		\newcommand\radd{\color{black}}
		\newcommand{\rremove}[1]{\ignorespaces}
\begin{document}

\title{GPU4S: Embedded GPUs in Space \\ Latest Project Updates}
\author{
	\IEEEauthorblockN{Leonidas Kosmidis\IEEEauthorrefmark{1},~~~~Iv\'{a}n Rodriguez\IEEEauthorrefmark{1}, ~~~~\'{A}lvaro Jover\IEEEauthorrefmark{1}, ~~~~Sergi Alcaide\IEEEauthorrefmark{1},	~~~~J\'{e}r\^{o}me Lachaize\IEEEauthorrefmark{2}}
\IEEEauthorblockN{~~~~~~Jaume Abella\IEEEauthorrefmark{1}, ~~~~~~Olivier Notebaert\IEEEauthorrefmark{2},~~~~~Francisco J. Cazorla\IEEEauthorrefmark{1}$^{,}$\IEEEauthorrefmark{3},~~~~David Steenari\IEEEauthorrefmark{4}~~~~}
\IEEEauthorblockN{}
\IEEEauthorblockA{
\IEEEauthorrefmark{1}Barcelona Supercomputing Center (BSC), Spain} \IEEEauthorblockA{\IEEEauthorrefmark{2}Airbus Defence and Space, France \\}
\IEEEauthorblockA{\IEEEauthorrefmark{3}Spanish National Research Council (IIIA-CSIC), Spain}
\IEEEauthorblockA{~~~~~~~\IEEEauthorrefmark{4}European Space Agency, The Netherlands
}
}

\maketitle

\begin{abstract}
Following the trend of other safety-critical industries like automotive and avionics,
the space domain is witnessing an increase in the on-board computing performance demands.
This raise in performance needs comes from both control and payload parts of the
spacecraft and calls for advanced electronics systems able to provide
high computational power under the constraints of the harsh space environment. On the non-technical side, for strategic reasons it is mandatory to get European independence on the used computing technology.
In this project, we study the applicability of embedded GPUs in space, which have shown a dramatic improvement of their performance per-watt ratio coming from their proliferation in consumer markets based on competitive European technology. To that end, we perform an analysis of the existing space application domains to identify which software domains can benefit from their use. Moreover, we survey the embedded GPU domain in order to assess whether embedded GPUs can provide the required
computational power and identify the challenges which need to be addressed for their adoption in space. In this paper, we describe the steps followed in the project, as well as a summary of results obtained from our analyses so far in the project.

%In the GPU4S project, we explore the possibility of using such devices in space. To that end, we survey the existing hardware and space application domains in order to identify potential hardware and software candidates, together with the challenges that need to be addressed for their adoption in space. In this paper, we describe the steps which will be followed in the project, as well as the results of our preliminary analysis in the first months of the project.

%The increasing performance demands in the space The aerospace domain is traditionally based on mature processing devices and custom hardware which satisfies the strict requirements of this environment. However, the existing solutions either cannot provide the increased computational needs of future space missions. The recent proliferation of embedded GPUs in consumer markets, which can provide high performance in a low-power envelope and most of which are designed in Europe, presents an unprecedented opportunity to the European space market to gain a competitive advantage while becoming self-dependent on these technologies. 

\end{abstract}

%\vspace{-0.1cm}
\section{Introduction and Background} 
\label{sec:intro}

The space domain needs high performance, scalable processing solutions for the increased computational requirements of future missions to enhance the autonomy and for on-board data processing.
Space presents a unique set of constraints with a small production volume compared to other safety critical industries{\radd{;}} therefore{\radd{,}} it is always keen to explore and adapt solutions from other domains in order to reduce non-recurrent costs if possible. The increased performance demand is required both for the \textit{platform computers}, which are responsible for controlling critical functionalities, like guidance, navigation an power distribution as well as the \textit{payload computers}, which are in charge of controlling the payload devices along with pre-processing of their data before they are transmitted to the ground. In the former case, more advanced capabilities are expected from future missions for i.e. increased autonomous navigation. In the latter case, the increase of the data generated by newer on-board scientific instruments which have higher resolutions and higher sampling frequencies requires more computational power to be accomplished.

Graphics processing units (GPUs) is a potential technology from another domain which can provide the higher performance needed. GPUs were originally used as an accelerator for visual applications, but nowadays they have evolved to general purpose high-performance accelerators, which in terms of performance and energy efficiency have surpassed the levels achieved by Central Processing Units (CPUs). The unprecedented level of performance per watt for very demanding computations led GPUs to become an integral part of high-performance computing. As a matter of fact, one quarter (132) of the supercomputers in the recent edition of the TOP500 list (as of November 2019) are based on GPUs, in addition to the Green500, which ranks the TOP500 supercomputers based on their performance per watt, 8 of the 10 top are using GPUs as accelerators for computing. For these promising reasons, past studies analysed the suitability of high-performance GPUs in space~\cite{hipnos}\cite{hipnos2}. However, those studies found that although their energy efficiency is high, their power consumption is an order of magnitude greater than the tight power budget of a space system, which is limited to a couple of Watts and for this reason are not suitable for this domain. The reason for this is not only because of the limited on-board power supply, but also because of the constraints in heat dissipation in the vacuum, which limits cooling solutions to passive dissipators only.

%Graphics processing units (GPUs), initially a special purpose type of accelerator for visualisation tasks, have since several years ago outperformed Central Processing Units’ (CPUs) raw performance and energy efficiency. This opens the door to achieve unprecedented performance with a very high energy efficiency for demanding computations, becoming essential for high-performance computing. As a matter of fact, one quarter (125) of the supercomputers in the recent edition of the TOP500 / Green500 list (as of June 2019) are based on GPUs, including the two most powerful supercomputers. 
%Past studies analysed the applicability of high-performance GPUs in space~\cite{hipnos}\cite{hipnos2}. Those studies concluded that although their energy
%efficiency is high, their power consumption is an order of magnitude higher than the limited power budget of a space system, which is limited to a couple of Watts. 

%, GPUs have entered in the embedded domain in order to satisfy the rising demand for multimedia
%multimedia-based handheld and consumer devices such as televisions, set-top boxes, in-vehicle entertainment and most notably in demanding games for smartphones.
Interestingly, GPUs entered in the embedded domain to satisfy the increasing demand for multimedia-based handheld and consumer devices such as smartphones, in-vehicle entertainment systems, televisions, set-top boxes etc. 
These embedded GPUs were re-designed, compared to the high-performance desktop/server ones, to exhibit low-power requirements targeting battery power and thermally-constrained devices.
Advances in transistor technology allowed these devices to achieve impressive performance that were only conceivable by high performance systems of the past decade~\cite{mobile_vs_console}.

%They were re-designed compared to their high-performance desktop/server counterparts to exhibit low-power requirements, essential for battery power and thermally-constrained devices. Improvements in the transistor technology allowed achieving impressive performance capabilities that were only possible in high performance systems of the past decade~\cite{mobile_vs_console}.

Despite the lower performance ratio between GPU vs CPU in the embedded domain than in the high-performance devices, due to power and thermal constraints, mobile GPUs were progressively adopted for accelerating heavy workloads, for applications ranging from signal processing, to advanced driving assistance systems (ADAS) in cars, as well as prototype supercomputers for exascale~\cite{montblanc}\cite{hpc}.

%Although the GPU vs CPU performance ratio is lower in the embedded domain than in the high-performance one due to power and thermal constraints, mobile GPUs are increasingly considered for accelerating heavy workloads, for applications ranging from signal processing, to advanced driving assistance systems (ADAS) in cars, as well as prototype supercomputers for exascale~\cite{montblanc}\cite{hpc}.

%These desired characteristics of GPUs make them a perfect candidate for computationally intensive tasks found in modern space systems. %For example,
%earth and deep space observation systems heavily build on image-processing,
%frequently accompanied with (lossless) compression to maximise the limited
%available downlink communication bandwidth. Moreover, future space applications
%increasingly include more advanced features. This includes fully autonomous
%guidance; navigation and control (GNC) based on computer vision; and machine
%learning both for i) earth orbit missions such as active debris removal, and ii)
%deep space exploration and interplanetary missions, including landing and rover
%navigation. This trend will significantly increase the demand for on-board
%processing capabilities, which existing general-purpose microprocessors such as
%LEON cannot cover.
%
Despite their promising characteristics, embedded GPUs have not been examined for their applicability in this domain. This project's goal is bridging this gap by providing an initial assessment of existing embedded GPUs, as a first step of their further exploration and adoption in this domain. {\radd{In particular, the project tries to address two questions: whether the algorithms used in on-board processing can be accelerated with GPUs, as well as whether embedded GPUs can provide the computational power required by future on-board processing, within the power limits of space systems. In addition, it is expected to scratch the surface regarding their reliability in the harsh space environment, so that it can be studied better in future projects, which will cover also other aspects that could not be completed in the limited time frame and scope of a pilot project.}}

%Despite their promising characteristics, embedded GPUs have not been explored for their applicability in this domain. This project aims at covering this gap by providing an initial assessment of existing embedded GPUs, as a first step of their further exploration and adoption in this domain. 

The rest of the paper is organised as follows.
Section~\ref{sec:pro} introduces the project and its main tasks. %, which includes the on-going and future activities. 
Section~\ref{sec:space_sw} analyses the performance demand of current and future space missions.
Section~\ref{sec:taxonomy} details the outcome of {\radd{the}} work on the analysis of the embedded GPU market.
Section~\ref{sec:results} presents the preliminary work performed so far, while Section~\ref{sec:future_work} provides the on-going and future work until the end of the project. 
{\radd{Section~\ref{sec:challenges} presents the challenges and difficulties experienced so far in the project.}}
Finally Section~\ref{sec:concl} presents the main conclusions of our work so far.

%\vspace{-0.1cm}
\section{The project}
\label{sec:pro}
{\rremove{This proposal}}{\radd{The GPU4S (GPU for Space) project, funded by the European Space Agency (ESA)}} aims to explore the suitability of embedded GPUs for space covering both the software and hardware perspectives. {\radd{The project is coordinated by the Barcelona Supercomputing Center (BSC), a leading institution in code parallelisation and optimisation, parallel programming models, critical systems and in particular embedded GPUs as well as an NVIDIA Center of Excellence. In addition it is home to two supercomputers in the actual TOP500 list, Marenostrum and Marenostrum P9 CTE. The latter is based on GPUs and it is currently in the 7th position of the most energy efficient supercomputers (Green500, November 2019). Airbus Defence and Space (ADS) in Toulouse, formerly Astrium, is Airbus' aerospace division, a world-leading primary spacecraft hardware and software supplier. The consortium partners have complementary expertise required to cover the wide range of activities required in the project.}}

\begin{figure}[t!]
				\vspace{0.2cm}
			  \centering
        \includegraphics[width=1\columnwidth]{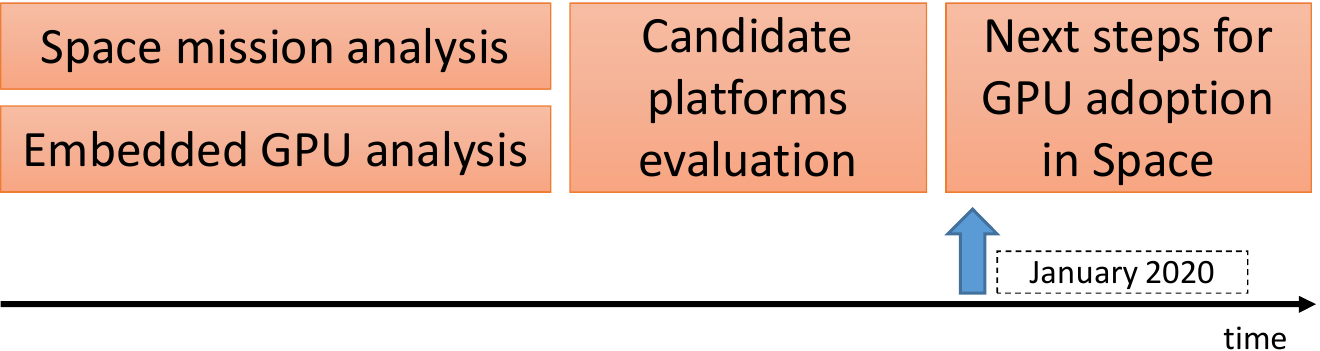}
        \caption{Main activities. Box width does not represent activity duration}
        \label{fig:timeline}
				%\vspace{-0.5cm}
\end{figure}

In order to reach its main goal, 
the project is organised in 4 main {\rremove{activities}}{\radd{work packages}} which are distributed in time as shown in Figure~\ref{fig:timeline}.

%The project explores the suitability of embedded GPUs for space from both, software and hardware perspectives. In order to reach its main goal, 
%the project is organised in 4 main activities as shown in Figure~\ref{fig:timeline}.

%\textbf{Space mission analysis}: In this party, we survey potential space areas, focus on in specific algorithms and applications that can take advantage form the use of GPUs in space, assuring that the selected applications are suited

\textbf{Space mission analysis}: {\rremove{In}}{\radd{in}} this part, {\radd{led by ADS,}} we study various space domains, targeting specific algorithms and applications that can benefit from the use of GPUs in space{\rremove{,}}{\radd{. BSC assists in this task by}} ensuring that the selected applications are suited for the GPU execution model.
This is needed to take into account because there are crucial design differences between GPUs and CPUs which are mandatory to be taken into account to understand how certain algorithms used in space, e.g. applications with divergent path execution among different threads or the ones that {\radd{do}} not use coalesced memory accesses, perform differently in embedded GPUs. In particular, certain space algorithms may not be suitable for embedded GPUs despite their high computational nature, and might require to be redesigned. 

%\textbf{Space mission analysis}: In this activity, we survey potential space areas and particular algorithms and applications that can benefit from the use of GPUs in space,
%ensuring that the identified applications are suitable for the GPU execution model. 
%This is important because there are fundamental design differences between GPUs and CPUs which are required to be taken into account to understand how certain algorithms used in space, e.g. those with divergent path execution among different threads or those not performing coalesced memory accesses, behave differently in embedded GPUs w.r.t. high-performance GPUs. In particular, certain space algorithms may not be suitable for embedded GPUs despite their high computational nature, and might require to be redesigned.

We provide a preliminary list of applications already validated as GPU-compatible in Section~\ref{sec:space_sw}. Furthermore, we characterize a list of criteria that can be
used for the selection of {\rremove{a}} GPU candidates depending on the mission profile.

%We provide a preliminary list of applications already validated as GPU-compatible in Section~\ref{sec:space_sw}. Moreover, we define a preliminary set of criteria that can be used for the selection of a GPU candidates depending on the mission profile.

\textbf{Embedded GPU analysis}: {\rremove{In}}{\radd{in this work package, which is led by BSC, in}} correlation with the previous activity, we review the available commercial-of-the-shelf (COTS) hardware and soft-IP (Intellectual Property) in terms of embedded GPUs{\rremove{,}}{\radd{. Our}}{\rremove{with the aim of}} {\radd{aim is to}} identify their characteristics, in order to select a candidate set of boards for evaluation in the next activities {\radd{based on a set of requirements commonly defined with ADS}}. We focus primarily on European IP, which leads the market and can provide complete autonomy in the European space sector. To understand the broad spectrum of mobile GPUs, we conduct a taxonomy of existing products.

%\textbf{Embedded GPU analysis}: In parallel to previous activity, we survey the available commercial-of-the-shelf (COTS) hardware and soft-IP (Intellectual Property), in order to identify their characteristics with the goal of selecting a candidate board for evaluation in the next activities. We  mainly focus on European IP, which in turn dominates the market and can provide complete independence in the European space sector. In this line, in order to understand the wide range of embedded GPUs, we perform a taxonomy of existing products. 

Besides the mission profile criteria, an important consideration when selecting a candidate board is energy performance and efficiency, because of the critical architectural distinctions between their designs and those of their high-performance counterparts, which must be exploited by the software. This fact has been studied only superficially so far \cite{date17}, and many works in the field have overlooked these differences and treated them as equivalents. Another aspect that is no less important is the available software ecosystem, which is vital for its application in space. Furthermore, we identify software applications in the space domain that can take advantage of GPUs, also justifying their use. Finally, it is very important to ensure whether these applications and their associated algorithms comply with the GPU programming model. 

%In addition to the mission profile criteria from the Space mission analysis, an important aspect for the selection of a candidate board, is the performance and energy-efficiency characteristics, due to the fundamental architectural differences between their designs and that of their high-performance counterparts, which need to be exploited by the software. This fact has only been superficially explored up to date \cite{date17}, with many works in the literature overlooking these differences and treating them as equivalent. Another equally important aspect is the available software ecosystem that is crucial for the applicability in space. Moreover, we identify software applications in the space domain that can benefit from GPUs and justify their use. Furthermore, it is important to ensure that these applications and their corresponding algorithms fit the programming model of GPUs. 

Then, we study the potential market options to select the suitable candidates for evaluation. We begin by providing a taxonomy of available GPU choices by listing the features and differences between products from various GPU suppliers or families for each category, i.e., GPUs and soft-IP. 
%
%Next, we survey the potential market options in order to select the appropriate candidates for evaluation. We started by providing a taxonomy of the available GPU options, listing the characteristics and the differences between products from different vendors or families of GPUs in each category, namely GPUs and soft-IP. 
%
The taxonomy and summary of our findings from the survey are discussed in Section~\ref{sec:taxonomy}.
	
%The classification and the summary of our preliminary observations of the survey are presented in Section~\ref{sec:taxonomy}. 

\textbf{Candidate Platform Comparison}: 
{\rremove{Based}}{\radd{based}} on the two initial steps, the identification of potential GPU candidate platforms follows. Then a comprehensive comparison among the selected embedded GPU candidates against current and upcoming processing devices for the space domain, enables to assess the potential benefits of embedded GPUs, particularly with regard to high performance requirements for future missions, while meeting the power and temperature limitations of the space environment.
%From both previous activities, we will identify promising GPU candidate platforms. As a next step we will perform a thorough comparison between the selected embedded GPU candidates and existing and future processing devices for the space domain. This way, the potential benefits of embedded GPUs can be evaluated, specifically regarding the high-performance requirements of future missions, while respecting the power and thermal limitations of the space environment.
%
For this{\radd{,}} we use representative kernels extracted from existing space algorithms to the GPU. Additionally, performance and other data from previous mission and application-specific integrated circuit (ASIC) processes will be used for comparison by standardizing them in accordance to the current space technology node (65nm). This will allow us to select the most adequate GPU platform for space from the evaluated candidates. {\radd{BSC is responsible for the benchmarking of the selected candidates, while ADS takes care of the benchmarking of existing space processors for the comparison.}}

%In order to do this, we will %define the evaluation criteria and 
%perform the evaluation by porting representative kernels derived from existing space algorithms
%to the GPU.  Moreover, performance and other data from past missions and previous Application-Specific Integrated Circuit (ASIC) processes will be used for comparison, by normalising them according to the current space technology node (65nm). This will allow the selection of the most appropriate GPU platform for space from the evaluated candidates.

\textbf{Next Steps for GPU adoption}: {\rremove{The}}{\radd{the}} final step of the study is to identify future steps in the adoption of GPUs in space by defining existing constraints as well as suggesting suitable solutions to address them. To determine the next stages of GPU adoption in the space domain with a system integrator approach, we are looking at the necessary steps for qualification of COTS systems by tackling their system-level reliability issues or the development of radiation-hardened components. In the first case, we are also delivering fault tolerance software solutions tailored specifically to these platforms, for the use of COTS embedded GPUs in space. Moreover, we port an algorithm provided by ESA to the GPU platform. {\radd{ADS is responsible for defining the adoption roadmap and the use of existing reliability solutions from its experience with designing space systems, while BSC provides reliability solutions particularly developed for GPUs, as well as the implementation of the GPU demonstrator.}}

%As final step of the project,  we will define the future steps required towards the adoption of GPUs in space, by identifying current limitations and proposing appropriate solutions to overcome them. In order to assess the next steps for the adoption of GPUs in the space domain with a system integrator view, we will examine the necessary steps for qualification of COTS systems by addressing their reliability concerns at system level or the development of radiation hardened components. In the former case we will also provide software fault-tolerance solutions specifically designed for these platforms, in case a COTS embedded GPU is selected in the previous task. Finally, an ESA provided algorithm will be ported to the GPU platform.

\section{Space Application Survey}
\label{sec:space_sw}

This section examines the domain of space applications in terms of the use of embedded GPUs.
%In this Section, we examine the space applications domain regarding the use of embedded GPUs.
%
The efficiency demands of space missions steadily increase. To illustrate the \textbf{current missions}, the Gaia astrometry mission~\cite{gaia} (dedicated to measuring the positions of celestial bodies), launched in 2013 by the European Space Agency as a follow-up mission to the Hipparcos mission~\cite{hipparcos} in 1989, is 100 times more accurate than its predecessor and is intended to map 1.7 billion stars, 4 times more than Hipparcos. In addition, these scientific missions produce vast amounts of data, which need to be transferred. Transmitting this amount of information is challenging even for our current communication standards. The majority of scientific on-board instruments, including those used for monitoring, use sensor arrays that enable parallel computing. For instance, the International Space Station's (ISS) alpha magnetic spectrometer (AMS-02) generates data at 7 Gigabit/s~\cite{ams}. Therefore, an array of 600 CPUs is used to reduce the amount of data by 3 orders of magnitude before the transmission~\cite{ams2}.

%The performance requirements of space missions are constantly increasing. As an example of \textbf{current missions}, the Gaia astrometry mission (devoted to the measurement of the positions of celestial bodies), launched in 2013 by the European Space Agency as a follow-up mission to the Hipparcos mission in 1989, has a hundred times better accuracy and aims to map 1.7 billion stars, 4 orders of magnitude more stars than its predecessor. Additionally, scientific missions can generate a massive amount of data, which may be challenging to transmit even on earth's surface. Most on-board scientific instruments, including the ones used for observation, use arrays of sensors, which allow for parallel processing. For example, the Alpha Magnetic Spectrometer (AMS-02) instrument at the International Space Station (ISS)~\cite{ams}, produces data in the rate of 7 Gigabit/s. For this reason, it uses an array of 600 CPUs to reduce the amount of data by 3 orders of magnitude, before transmission~\cite{ams2}.

Lastly, upcoming space missions, which include advanced space concepts like space-tug and the Active Debris Removal (ADR) system, will require major computational power to implement new features, specifically for autonomous guidance and navigation control (GNC) based on image processing and autonomous learning~\cite{adr}, to identify, approach, gather and eventually clear debris. 

Likewise, on-board data processing for new generation missions will grow in almost all mission types~\cite{dasia12}\cite{eucass15}\cite{amicsa16}:

%Finally, future space missions such as the Active Debris Removal (ADR) will need an unparallelled amount of computational power to achieve new functionalities, for autonomous guidance and navigation control (GNC) based on image processing and machine learning~\cite{adr}, in order to identify, approach, rendez-vous and finally remove the target. 
%
%
%Similarly, the on board data processing for next-generation missions will increase in almost every kind of mission~\cite{dasia12}\cite{eucass15}\cite{amicsa16}:

\begin{itemize}
\item For robotic exploration and research, high-performance data collection networks are required to satisfy the heavy data rate coming from on-board equipment (spectrometers, imagers, etc.).
%\item For Science and robotic exploration, high data performance acquisition chains will be required to meet the high data rate generated by instruments (spectrometers, imagers etc.) as the importance of on board-autonomy and processing needs for planetary exploration is increased.

\item For Earth monitoring, the increase in sensor technology resolution, support for dynamic range and faster read rates have resulted in a drastic growth in sensor bandwidth and data volume, causing significant downlink capacity bottlenecks as well as creating a demand for very high on-board data processing capabilities, for both data processing and compression. Furthermore, emerging technologies, such as deforming mirrors, need significant on board processing. Streaming video recorded from space could also be a potential new application either instead of static observation images or as flight mission data for postmortem analysis. Moreover, image enhancement is another possible direction. 
%For Earth Observation, the evolution in sensor resolution, dynamic range and faster readout rates has led to a dramatic increase of sensor bandwidth and data volume, creating significant bottlenecks in downlink capacity and a need of very high on-board data processing capabilities for both data processing and compression. Moreover, new technologies such as deforming mirrors, require important on-board processing. Video from space could also be a new application, as well as the image enhancement based on the acquisition of several images. 

\item For next-generation launchers, the considerable increase in data speed and on-board computing power will allow worthwhile applications, like the increase in user telemetry transmission channels. Among them, video data transmission for monitoring agile and critical launch maneuvers, currently not feasible with data transfer rates of up to 400 Mbits/s at a compression ratio of 20.
%\item For next generation launchers, considerably increasing data rate and on-board processing capabilities can enable interesting applications, such as increased user telemetry transmission channels. This includes the transmission of video data to monitor critical agile launch maneuvers, which are not possible with today’s transmission rates of up to 400 Mbits/s with a compression ratio of 20.

\item The navigation equipment for future missions, such as the space tug concept for the Active Debris Removal (ADR) project, which can shift the orbit of non-cooperative spacecraft, will rely on computer vision. 
%\item The navigation equipment for new missions like the space tug concept required for Active Debris Removal (ADR)~\cite{adr} which will be able to change the orbit of non-cooperative spacecraft, will be based on computer vision. This type of processing, including localisation and mapping, is essential and very demanding. 

\item In the telecommunications field, upcoming developments go beyond typical geosynchronous equatorial orbit (GEO) based telecommunication systems, including missions such as machine-to-machine communication from low-Earth orbit with medium to large spacecraft constellations or spectrum monitoring.
%\item In the telecommunication domain, future applications go beyond the classical geosynchronous equatorial orbit (GEO) based telecommunication systems, and include missions such as spectrum surveillance or machine to machine communication from the Low Earth Orbit with medium or large spacecraft constellations.

\item It is also expected that radar processing will undergo a similar technological evolution and will be synergistic with the telecommunications domain, as both demand high-performance state-of-the-art signal processing.
% Radar processing is also expected to follow a technological evolution similar to and synergistic with the telecommunication domain, since both require high-performance signal processing.

\item The only domain where performance requirements are not expected to increase is launchers. In the case of scalable launchers, high performance processing requirements are not rapidly evolving, as the need for in-flight video transmission (mainly for post-mortem analysis) requires a strong data compression ratio: in the order of 20. This is already met and the need is not expected to evolve rapidly, remaining in a range similar to that of current Earth Observation needs. Nevertheless, with reusable launchers, the re-entry/landing phases alone require high-availability, high-performance processing devices, especially for vision-based positioning and landing guidance. For this study, such application is considered similar to that of Planetary Approach and Landing already covered by the exploration domains. Therefore, the launcher domain is not considered as candidate for GPU acceleration in this project.

%\item Launchers is the only domain in which performance requirements are not expected to grow. For expandable launchers, requirements for high performance processing are not growing very fast since the need for video transmission during the flight (mainly for post-mortem analysis) requires a strong data reduction ratio: in order of 20. This is currently covered and no fast evolution of the need is foreseen, remaining in a similar range to the current Earth Observation requirement. On reusable launchers, the re-entry/landing phases will require high performance highly available processing devices especially for vision based navigation and guidance for landing. In this study, such application is considered as similar with the Planetary Approach and Landing already covered by the exploration domain. So the launcher use case will not be particularly considered further in this study.

\item Last but not least, new missions will need considerable agility and autonomy provided by sophisticated robotics for in-orbit operation and space exploration, from robotic navigation to landing. Also, the use of artificial intelligence applications is expected to increase in on-board devices due to the high complexity of some space exploration tasks.
%\item Finally, future missions will require significant autonomy and agility provided by advanced robotics for in-orbit operation and space exploration, including landing and rover navigation. In the long term, it is anticipated that increasing artificial intelligence will be required on-board most spacecraft.
\end{itemize}
\vspace{-0.1cm}
 
%Those high processing needs have as a consequence to look for space computing
%solutions among available technologies such as general purpose processors, DSPs,
%FPGA, ASIC etc.

The development of complex computing architectures embedding many processing cores on single device with low power consumption as GPUs is a real opportunity to significantly enhance on-board processing power and meet the performance needs of future space applications.
%The emergence of %complex computing architectures embedding many processing cores on single device with low power consumption as GPUs is a real
%embedded GPUs with high performance and low power consumption presents a real
%opportunity to drastically increase the on-board processing capability and
%satisfy the future space applications' performance needs.
\begin{table}[h!]
%\vspace{-0.4cm}
\center
\caption{\label{tab:alg}Representative spatial domains and algorithms suitable for GPU acceleration.}
\includegraphics[width=1.0\columnwidth]{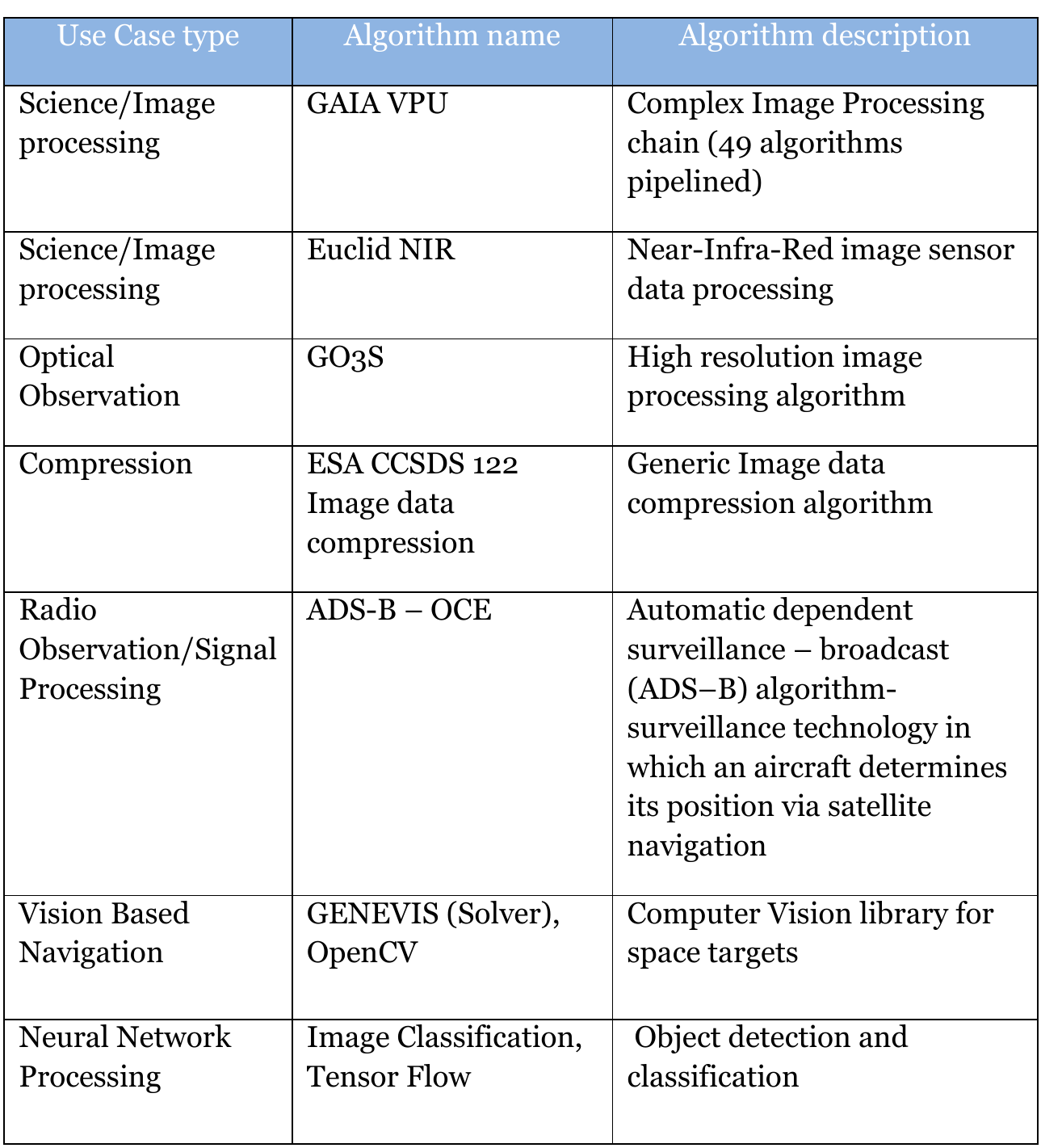}
\vspace{-0.9cm}
\end{table}

We have selected a first set of applications that fit into the areas identified above and are presented in the Table~\ref{tab:alg}. Some of these highly computationally intensive algorithms have already been chosen for the processing of high quality data in present and upcoming missions. We believe that all of the representative algorithms in the table may be well suited for GPU acceleration. The reason for this is that the majority of GPU microarchitectures experience a severe throughput penalty when software behaves unpredictably, such as not accessing consecutive memory locations (memory coalescing) as well as taking different paths (branch divergence). Preliminary studies of these applications indicate that they are free of branch divergences, while their memory is accessed in a regular way.

%A first set of applications fitting in the above-identified areas has been selected and presented in the Table~\ref{tab:alg}. Several of these computationally demanding algorithms have already been selected for high performance data processing in current and future missions. Our analysis indicates that all the representative algorithms in the table could be appropriate for GPU acceleration. This is based on the fact that most GPU microarchitectures suffer a significant performance penalty with software exhibiting irregular behaviour, e.g. not accessing consecutive memory positions (memory coalescing) or taking different paths (branch divergence). A first study of those applications shows that they are free of path divergence, while their memory accesses are regular.

The knowledge already gained and the leveraging of those available and operational software extracts in past and recent parallel studies ensure both a good foundation for this GPU work and a potential starting point for performance and porting effort benchmarking. To finalize the survey and guarantee that as many possible domains are covered, an internal meeting will be organized with all Airbus programs to pinpoint other algorithms that may be attractive for GPU implementation.
%Already acquired knowledge and exploitation of those existing and operational software extracts in previous and current parallel studies guarantee both a good starting point for this GPU work and, potentially, an existing baseline for performance and porting effort comparisons. In order to complete the survey and ensure to cover as many potential domains as possible, an internal meeting with all Airbus programs will be organised to identify further algorithms that could be interesting for GPU implementation.

The study of space application domains also involves defining a set of criteria for eligible GPU candidates, supported by the specific features of certain application and mission domains. As an example, Low Earth Orbit (LEO) missions have lower availability requirements that permit the use of COTS components, particularly if such components have reliability characteristics as pre-qualified products for use in industrial or safety-critical domains.
%The space application domain survey includes also the definition of a set of criteria for the selected GPU candidates based on the specific details of particular application domains and missions. For example, Low Earth Orbit (LEO) missions have less stringent availability requirements which allow the use of COTS components, especially if they have reliability features as products that are already qualified for use in industrial or safety critical domains such as the automotive.

Some of the parameters that we have identified from the mission profile are: the long term availability of the technology/equipment to match the long lifespan of space missions, the feasibility of the implementation of the algorithms on the GPU, the power and thermal limitations of the mission, the I/O interfaces, the one-time cost of developing radiation tests and mitigation technologies, the recurrent expense for the number of pieces of equipment, in addition to the flexibility to use various algorithms at specific stages of the mission schedule and the update of the algorithm for correction or extension.
%Some of the metrics we have identified so far coming from the mission profile include: the availability, usable in space, of the technology/equipment, the algorithms implementation feasibility, the power and thermal constraint of the mission, the I/O interfaces, the non-recurring development cost for radiation testing and the development of mitigation techniques, the recurring cost of the equipment regarding the number of pieces, and the flexibility to use different algorithms at different steps of the mission timeline and the update of the algorithm for correction or extension.

\section{GPU Taxonomy}
\label{sec:taxonomy}

\begin{figure}[h]
\centering

\includegraphics[width=0.8\columnwidth]{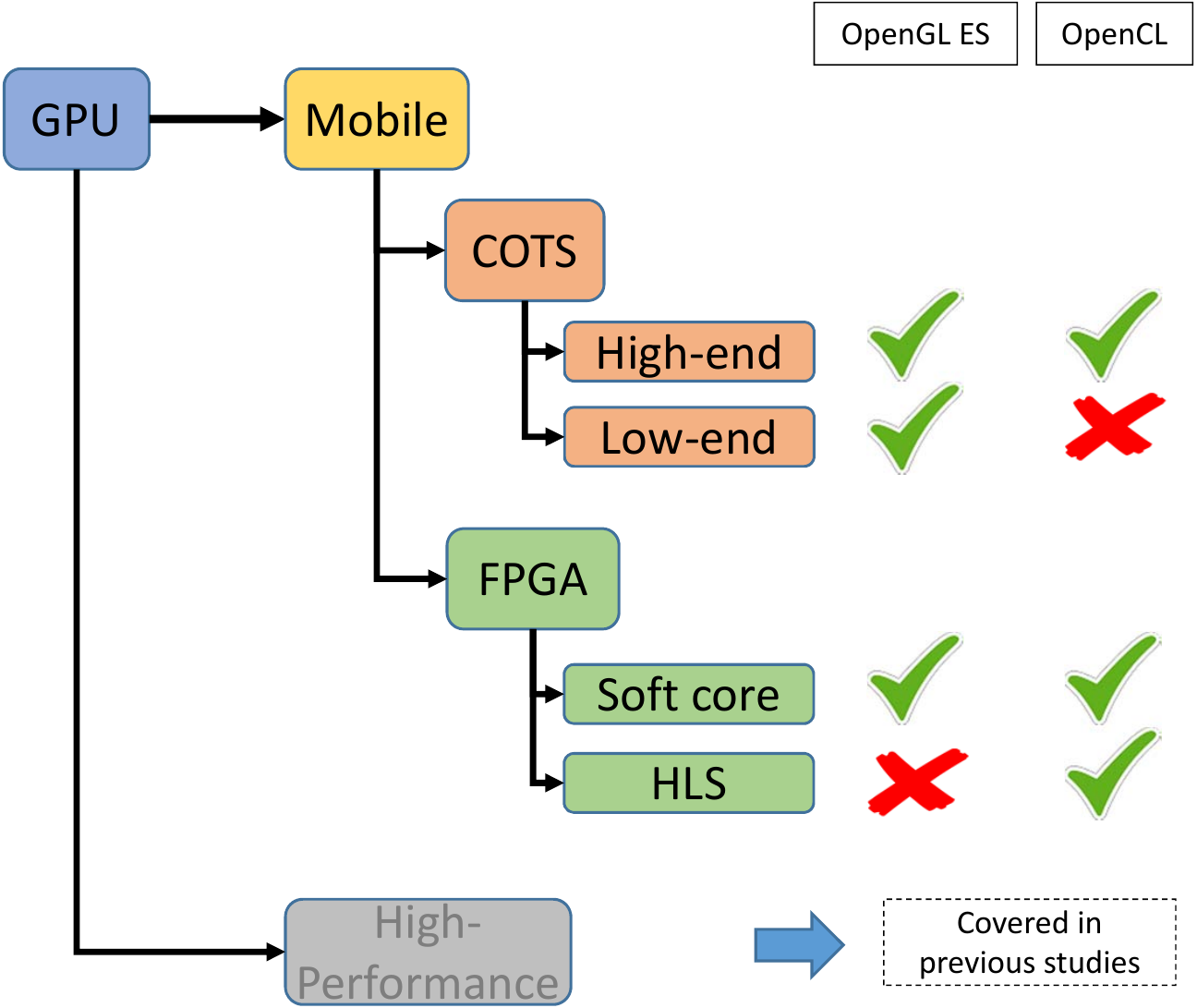}
\caption{\label{fig:taxpic}GPU Taxonomy}
%\vspace{0.5cm}
\end{figure}

This section presents a classification of mobile GPUs, shown in Fig~\ref{fig:taxpic}, in order to elaborate on the different potential options available for the space domain. 

%In this Section, we present a taxonomy of mobile GPUs, summarised in Fig~\ref{fig:taxpic} in order to better explain the various options of potential candidates for the space domain. 

The first category classifies GPUs as embedded or high-performance. This project only focuses on the embedded ones, since the high-performance ones were already covered in previous works~\cite{hipnos}\cite{hipnos2}.

%At the top level we classify GPUs as either embedded or high-performance and focus just on the former category, since the latter was covered in previous studies~\cite{hipnos}\cite{hipnos2}.

Unlike the desktop GPU market where 3 major players dominate (NVIDIA, AMD and Intel), the embedded GPU market has many major vendors, among which Europe stands out as a major force in which some of the most important embedded GPU IP companies began to emerge: ARM, Imagination Technologies or Bitboys, between others. The latter started the development of Imageon, now re-branded to Adreno by Qualcomm in Finland, after being acquired by ATI and AMD. In a similar way, the Broadcom's VideoCore of the European Raspberry Pi educational computer was initially developed by Alphamosaic (UK). In addition, ThinkSilicon, which specialises in GPUs for IoT devices, is based in Greece.

%The first difference we observed in the mobile GPU market is that, unlike the desktop GPU market which is dominated by 3 major players (NVIDIA, AMD and in lesser degree Intel), the embedded GPU market features several big GPU IP providers. Most of them are European (ARM, Imagination Technologies) or started as European (Adreno was initially designed under the Imageon brand name by the Finnish company BitBoys, now Qualcomm Finland, after acquisition by ATI and AMD). Similarly Broadcom's VideoCore used in the low-cost European educational computer Raspberry Pi has been developed by the UK-based Alphamosaic before it was purchased by Broadcom.
%In addition, ThinkSilicon which specialises on GPUs for IoT devices is based in Greece.

Another relevant point of our work is that thanks to previous works~\cite{date17} \cite{date16}\cite{brook_auto}, we were able to broaden the scope of the study past the evident choice of OpenCL supported devices (see Fig~\ref{fig:taxpic}), enabling wider market coverage of existing embedded GPUs while also providing a wider range of choices for the right choice of embedded GPUs. It is important to note that the completion of the survey, particularly with regard to the potential IP acquisition of embedded GPUs to manufacture a rad-hard chip, requires contacting various suppliers, which translates into a lengthy administrative process that is underway.

%Another key element in our work is that based on previous works~\cite{date17}\cite{date16}\cite{brook_auto} we were able to extend the survey beyond the obvious choice of OpenCL programmable devices (right part in Fig~\ref{fig:taxpic}), which allows better covering the available embedded GPU market and providing a wider spectrum for the appropriate selection of embedded GPUs. Note that in order to complete the survey, especially regarding the potential embedded GPU IP acquisition to make a rad-hard chip, it is required to contact
%access non-public information from 
%the various vendors, which requires a long administrative process that it is currently under its way.

Mobile GPUs can be split into two categories: FPGA implementations and COTS GPUs. In the first category, FPGA implementations support circuit reconfiguration to ease hardware development. This is useful when a product is not completely specified or additional needs are introduced at a later stage. In the latter category we find mobile GPUs that are implemented as ASICs (Application-Specific Integrated Circuits), usually as part of a larger SoC (System-On-Chip) that often contains CPUs and other devices such as embedded memory, DSPs, etc. Both of these categories are interesting for the space domain.

%Mobile GPUs can be first divided into two broad categories, COTS GPUs and FPGA implementations. In the former category we find mobile GPUs which are implemented as ASICs (Application-Specific Integrated Circuits), typically as part of a larger SoC (System-On-Chip) which usually contains CPUs and other devices e.g. DSPs, embedded memory etc. In the latter category, FPGA implementations allow reconfiguration of the circuitry in order to facilitate hardware development, especially when a product is not fully defined or additional requirements are introduced at a later time. Both categories are interesting for the space domain. 

\subsection{COTS GPUs}
%\vspace{-0.2cm}

COTS designs deliver low recurring costs and improved efficiency compared to FPGA implementations using the same design methodology. COTS components are made using the latest generation of advanced silicon technologies, such as 10nm FinFET, as they are directed at mainstream consumer devices. However, the reliability in these markets with respect to both temporary and permanent failures has not been tackled, as these devices are not used in critical systems and users are generally expected to replace them after every 2-3 years with newer ones with more advanced capabilities. Consequently, an extra effort is needed to protect these components from the effects of radiation on both software and hardware.

%COTS designs provide small recurring costs and higher performance compared to FPGA implementations using the same design technology (often specified in nm). COTS components are manufactured in advanced state-of-the-art silicon technologies like 10nm FinFET since they target widely used consumer devices in order to benefit from the higher performance and power characteristics of those processes. However, the reliability in those markets regarding transient and permanent faults has not been addressed, since these products are not used in critical systems and they are expected to be replaced voluntarily by the users every 2-3 years with later products with more advanced capabilities. Therefore, additional effort is required in order to shield those components from radiation effects either in hardware or in software.
 
The COTS Embedded GPU category can be divided into Low-End and High-End products. 

\vspace{0.2cm}\noindent\textit{A.1. Low-end GPUs}
\vspace{0.2cm}

Low-end GPUs only support graphics APIs such as OpenGL ES 2 and hold a large niche in the mobile industry~\cite{gpu_market_2}\cite{gpu_market_3}. The simplified architectural design of these products makes the support of OpenCL impossible, however, thanks to this design approach, the power consumption is considerably reduced. ARM's Mali-4XX is the most significant example,  which accounts for 20\% of the mobile phone sector's market share~\cite{mali400_market}, excluding other widely produced mass-market devices such as single board computers, set-top-boxes, TV-sets, automotive systems and FPGAs. 
%
%Low-end GPUs support only graphics APIs such as OpenGL ES 2 and have a large share of the mobile market~\cite{gpu_market_2}\cite{gpu_market_3}. These products feature a simpler architectural design, which is incapable of supporting OpenCL but results in lower power consumption. The most prominent example is ARM's Mali-4XX, probably the most licensed embedded GPU to date, which has a 20\% of the smartphone market share~\cite{mali400_market}, without including in this market portion other
%massively produced consumer devices such as single-board-computers, set-top-boxes, tv-sets, automotive systems and FPGAs, such as the Zynq UltraScale+. 
%
An advantage of such a long-standing and widely used device is the maturity of the technology and the development tools accompanied usually by a well defined documentation.

%An advantage of such a widely used older device is the maturity of technology
%and its associated development tools, which provide evidence from use and
%documentation of their known problems, if not already fixed in later
%hardware/driver revisions.

Other devices from this category are Qualcomm's Adreno 2xx, Broadcomm's VideoCore IV, Imagination Technologies PowerVR SGX and ThinkSilicon's NemaPico and NemaTiny. While OpenCL support is not provided on these low-end devices, it is possible to program efficient solutions for general purpose computations employing general purpose computations over embedded graphics~\cite{date16}\cite{brook_auto}.

%Other products in this category from different vendors are Qualcomm's Adreno 2xx, Broadcomm's VideoCore IV, Imagination Technologies PowerVR SGX and ThinkSilicon's NemaPico and NemaTiny. Although OpenCL support is not provided on low-end GPUs, and cannot be implemented without low-level access to the GPU design, efficient solutions for general purpose computations on them exist~\cite{date16}\cite{brook_auto}.

One of the main issues of both high and low-end embedded GPUs is the lack of public domain architectural designs, which prevents researchers to certainly know the full execution pipeline of a device under a specific program execution, which also does not allow observability for space qualification. Vendors only supply high-level implementation details and all of the development tools are proprietary and close source. The only GPU with open specifications to date is Broadcom's VideoCore IV, but the limited number of development tools available work at the assembly level. This results in low productivity and high complexity, while no debugging or profiling methods are available.

%A potential issue with the adoption in space of both low and high-end embedded GPUs, is the non-disclosed architectural design of almost all embedded GPU, which prevents getting the required full control over the execution and observability for space qualification. The vendors provide only high-level information about the implementation and all the development tools are closed source. The only GPU with open specification so far is the VideoCore IV from Broadcom, however the limited available development tools from the open source community work at assembly level. This translates to low productivity and high complexity, while there is no debugging or profiling method available.

% vvvvvvvvvvvvvvvvvvvvvvvvvvvv NO vvvvvvvvvvvvvvvvvvvvvvvvvvvvvvvvvvv
%At first iteration, the Xilinx Zynq Ultrascale+ featuring a multicore ARM Mali-400 GPU appears to be a very good candidate of a COTS embedded GPU if this technology is chosen. The OpenGL ES 2 compliant ARM Mali400 multicore GPU for
%embedded systems could target aerospace processing applications, providing significant computing power which can be leveraged by BSC’s technology for GPGPU programming over mobile graphics~\cite{brook_auto}. Moreover, the platform
%features embedded safety mechanisms.

\vspace{0.2cm}\noindent\textit{A.2. High-end GPUs}
\vspace{0.2cm}

High-end embedded GPUs support both graphics and computing APIs, such as OpenCL or NVIDIA's proprietary GPU programming model, CUDA. These architectures are the ARM Mali T6xx-T8xx and G7X families, the latest SGX and Rogue families from Imagination, Adreno 3xx and above from Qualcomm, NemaSmall from ThinkSilicon and the latest Vivante GC series. While these architectures may, in principle, have the ability to support OpenCL runtime, GPU providers do not always release drivers. In fact, Google recently dropped OpenCL use in Android for this same reason. As a result, GPU manufacturers are less interested in developing OpenCL drivers for mobile GPUs unless there is a clear interest from large companies or other sectors, such as supercomputing. On the other hand, although NVIDIA supports OpenCL in its high-performance GPUs, in its embedded GPUs like the Jetson Series, composed by K1, X1, X2 and Xavier, only supports CUDA. Thus, the choice of a high-end GPU has to entail a deeper analysis than just the review of the providers' product sheets.

%GPUs in the high-end of the embedded spectrum support both graphics and compute APIs, such as OpenCL. Such architectures are ARM’s Mali T6xx-T8xx and G7X families, Imagination’s latest SGX and Rogue families, Qualcomm's Adreno 3xx and above, ThinkSilicon's NemaSmall and Vivante's latest GC series. Despite those architectures theoretically can support an OpenCL runtime, GPU vendors do not always provide a driver. In fact, OpenCL is usually only available to developers for use with certain development kits, while recently Google dropped its use in Android. As a consequence, GPU vendors are less keen to develop, release or support OpenCL drivers for mobile GPUs, unless there is an explicit interest from large companies such as Samsung or Sony  for certain GPU models in their flagship products, or from other domains such as supercomputing. Therefore, the selection of a high-end GPU has to involve a deeper analysis than a simple analysis of vendors' products sheets. 

As the majority of embedded GPUs are primarily aimed at consumer markets, they do not explicitly comply with safety requirements. The exceptions here are Imagination's PowerVR 6XT (GX6650) GPU, which is found in ASIL-B certified automotive platforms like the R-Car H3 from Renesas, and the latest series of Furian GPUs from Imagination and Xavier from NVIDIA, which are designed to be ASIL-D certified, the highest level of safety for automotive applications.

%Since most of the embedded GPUs target mainly consumer markets, they don't address explicitly safety requirements. The exceptions in this is case are Imagination's PowerVR 6XT (GX6650) GPU found in ASIL-B certified automotive platforms such as the Renesas R-Car H3 and Imagination's latest GPU Series Furian and NVIDIA's Xavier, which are designed targeting ASIL-D certification, the highest assurance level in automotive systems.

{\radd{
\vspace{0.2cm}\noindent\textit{A.3. Machine Learning and other Accelerators}
\vspace{0.2cm}

In the previous section, we identified the increased autonomy as an emerging on-board requirement, which requires neural network processing to be implemented. These algorithms are very computationally demanding and GPUs have been identified as suitable architectures for their acceleration. However, almost all embedded GPU design companies have introduced their own machine learning accelerators together with custom software stacks, which can provide higher performance neural network processing, mainly inference, in a more energy efficient way. Similarly, they introduce other special purpose accelerators eg. for vision workloads.

For example, the Project Trillium from ARM includes the Machine Learning Processor, ARM's Object Detection processor, as well as ARM's NN (Neural Network) software. Moreover, ARM's neural network software allows interoperability of machine learning operations on the CPU, GPU or the specific accelerators that can be found in their SoC. Similarly, Imagination Techologies provide Series2NX, a family of neural network accelerators with variable low arithmetic precision from 16 bit down to 4 bits and a software stack with various machine learning frameworks.
Think Silicon offers the NEMA xNN power efficient inference accelerator which features 8 bit operations as well as approximate computations.
Finally, Nvidia includes in its Xavier SoC Tensor Cores, which is a matrix multiplication accelerator for machine learning workloads, NVDLA Nvidia's Deep Learning Accelerator which is released as open source IP and can accelerate various deep learning algorithms, as well as the PVA (Programmable Vision Accelerator) an image processing VLIW-based (Very Long Instruction Word) accelerator. The latter is similar to Intel's Myriad X VPU (Vision Procesing Unit), another european developed technology from Modivious in Ireland, before it was acquired by Intel.

Each vendor provides its own software stack for these accelerators together with optimisation tools and support for popular machine learning frameworks like Tensorflow, Pytorch etc. However, usually these accelerators are tied to the GPU of the corresponding vendor. Moreover, the machine learning frameworks are constantly evolving due to the popularity of this domain. For these reasons the experimental evaluation and fair comparison of such accelerators is very complicated. Therefore it is considered out of the scope of our project and may be probably investigated in a future ESA activity. Instead, we focus on running the machine learning workloads on the embedded GPUs, which might not be as energy efficient as the NN accelerators, but they are quite efficient for such workloads. More importantly GPUs are more flexible than ASIC accelerators and they are programmed with more general purpose APIs.
}}

\subsection{FPGA Solutions}
%\vspace{-0.2cm}

FPGA solutions provide inferior throughput, however, the underlying COTS FPGA device can be radiation hardened and qualifiable to be used in space, such as the V5QV from Xilinx. Furthermore, research into emerging silicon technologies has demonstrated higher levels of reliability than existing technology used in space (65 nm), increasing expectations concerning the use of these technologies in space. 

%FPGA solutions offer lower performance but the underlying COTS FPGA device can be radiation hardened and qualified for space use such as Xilinx's high-density single-event immune V5QV. Moreover, studies of new silicon technologies have shown better reliability than the current technology used in space (65 nm), increasing the ambitions for use in space. 

The Xilinx report about Failures in Time (FIT) in the Xilinx FPGA~\cite{reliability} demonstrates a technology improvement which will be verified with the next generation of FPGA. These solutions can provide further benefits for long-term interplanetary missions. In particular, the configurability of the FPGA can shorten the time-to-launch of a several year mission, even if the desired hardware accelerator is not fully built. Alternatively, in the event that a new, more efficient image compression algorithm is developed, a hardware accelerator can be repurposed to support it, thereby decreasing the size of the data consequently reducing the time for downlink communications. For the aforementioned explanations, both of these potential approaches have their unique strengths, and are thus included in our study.

%The Xilinx report about Failures in Time (FIT) on Xilinx FPGA~\cite{reliability}, shows an improvement with technologies that shall be confirmed with the next FPGA generation. Such solutions can provide additional advantages for long interplanetary missions. In particular, the FPGA configurability can reduce the time-to-launch of a several year mission, even in the case that the desired hardware accelerator is not fully developed, debugged or tested. Or, in the case that a new more effective image compression algorithm is invented, a hardware accelerator can be reprogrammed to support it, thus reducing the data size and therefore the time for downlink communications. For the above reasons, both potential solutions have their unique advantages, and therefore they are considered in our survey.

This category can be subdivided in two levels:

%This group can be further divided into two categories:

\vspace{0.4cm}
\noindent\textit{B.1. Soft GPU cores}
\vspace{0.2cm}

Soft embedded GPU cores are implemented in RTL (Register Transfer Level) by using a hardware-description language such as VHDL or Verilog. The design is then synthesized into the FPGA, which can be used seamlessly by the software, either through graphics or compute APIs.

%Soft embedded GPU cores are implemented in RTL (Register Transfer Level) using a hardware description language such as Verilog or VHDL. The design is then synthesised on the FPGA, where it can be used transparently to the software either using graphics or compute APIs.

%The first category is composed by 
%Soft embedded GPU cores are implemented in RTL (Register Transfer Level) using a
%hardware description language such as Verilog or VHDL. The design is then
%synthesised on the FPGA, where it can be used transparently to the software
%either using graphics or compute APIs. %Note that most companies designing
%embedded COTS GPUs can in theory offer this possibility for licensing their IPs, 
%They do it because all these companies use FPGAs for their product development
%and, moreover, 
%since they are fabless, so they do not manufacture their own
%chips, but instead provide their IP in source or encrypted form in order to be
%integrated in a SoC and be manufactured. 

Certain GPU design companies offer evaluation solutions for specific FPGA devices, such as Think Silicon with its high-end NEMA GPU and the low-end Think2.5D products for Xilinx's Zynq platforms. But the feedback we have gathered from all the commercial GPU IP providers to date is that most high-end embedded GPUs outperform existing FPGAs, and only small configurations of these designs can be accommodated on very costly ($\sim$50,000 euros) FPGA development boards. Moreover, the IP providers claim that the achieved target frequencies of such designs on the FPGA are very low, which in combination with the reduced configuration of the designs in terms of cores and cache sizes in order to fit in the FPGA, results in very low performance in comparison to their ASIC implementation. Hence, we do not recommend the use of commercial GPU cores for FPGA implementations, but only for rad-hard ASICs.

%Some GPU designer firms explicitly offer evaluation solutions for certain FPGA devices such as Think Silicon with their high-end NEMA GPU and low-end Think2.5D products for Xilinx's Zynq platforms. However, the feedback we have received from all commercial GPU IP vendors so far is that most of the high-end embedded GPUs exceed the capacity of existing FPGAs and only reduced configurations of such designs can fit on very expensive ($\sim$50K euros) FPGA development boards. In addition, according to the vendors the achieved target frequencies of such designs on the FPGA are very low and in conjuction with the reduced configuration, they result in very low performance compared to their ASIC implementation. Therefore, we do not suggest using commercial GPU IP cores for FPGA implementation but only for rad-hard ASICs.

Open source research-oriented soft GPU cores are also available, like MIAW \cite{miaw} and FlexGrip \cite{flexgrip}, which implement AMD and NVIDIA as GPU microarchitectures or FGPU \cite{fgpu} which is different from any commercially available GPU. Unfortunately, these cores do not have energy efficient GPU microarchitectures such as tiled and deferred rendering architectures~\cite{date17}. These cores support only a subset of the instruction sets (typically limited to integer instructions) and offer support only for the compute APIs, but not graphics. On the other hand, these projects come with development tools of limited functionality, and without debugging or profiling capabilities, whereas most of them are no longer maintained or supported in any way. Aside these problems, the most significant obstacle to the use of these designs in space is their licensing terms.  Having a GPL license in most cases, would mean releasing the RTL code for the complete platform implemented in the FPGA. For this same reason, open source GPU designs are not appropriate for this domain.

%Open source research-oriented soft-GPU cores are also available, such as MIAW  \cite{miaw} and FlexGrip \cite{flexgrip}, which implement AMD and NVIDIA like GPU microarchitectures or FGPU~\cite{fgpu} which does not resemble any commercial GPU. However, these cores do not have a low-power GPU microarchitecture -- such as tile-based and deferred rendering architectures~\cite{date17} --, only implement a subset of the instruction sets (typically limited to integer instructions) and support only compute APIs, not graphics as well. Moreover, the projects come with limited-functionality development tools, and without debugging or profiling capabilities, while most of them are not maintained any more and therefore lack any type of support. Besides these problems, the most important roadblock for using these designs in space is their licensing conditions. Being GPL-licensed for the majority of the cases, using them in a commercial setup with proprietary hardware IP blocks such as SpaceWire, would require the release of the RTL code of the entire platform. For this reason, open source GPU designs are not suitable for this domain.
 
\vspace{0.2cm}\noindent\textit{B.2. High-Level Synthesis}
\vspace{0.2cm}

Modern FPGAs also support OpenCL using High-Level Synthesis (HLS). These products convert OpenCL into custom circuits, which are shaped to run in the FPGA network. While this is not a GPU solution, it is based on OpenCL, which delivers the same application interface as a high-end embedded GPU or a soft GPU core. All of these components can be found in both Xilinx and Altera, including Intel's recent HARP prototypes, with both a CPU and an FPGA on the same chip. This reconfigurable solution has the advantage that the FPGA hardware resources can be used more efficiently across multiple algorithms in comparison to a fixed soft-GPU design solution. However, FPGA reconfiguration (flashing) requires much more time than running different kernels on a GPU, and whilst FPGAs are now used in space missions, this feature has never been used before.

%Modern FPGAs also support OpenCL using High-Level Synthesis (HLS). Such products translate OpenCL to custom circuits, which are configured for execution on the FPGA fabric. Although this is not a GPU solution per se, this solution is based on OpenCL, which provides the same software interface as a high-end embedded GPU or soft-GPU core. Such products are both available from Xilinx and Altera, including recent Intel's HARP prototypes, which integrate both a CPU and an FPGA in the same chip. The reconfiguration of this solution provides the advantage that the hardware resources of the FPGA have the potential to be utilised more effectively among various algorithms compared to a fixed-design soft-GPU solution. However, FPGA reconfiguration (flashing) takes much longer than executing different kernels on a GPU, and although FPGAs are currently used in space missions, this functionality has never been used before.

High-level synthesis in OpenCL makes the development and debugging effort significantly easier versus a hardware description language. However, the performance of the generated circuits from the high-level synthesis needs to be evaluated.

%High-level synthesis in OpenCL facilitates significantly the development and debugging effort compared to a hardware description language. However, the efficiency of the generated circuitry from the high-level synthesis hast to be evaluated.

In addition, existing space-qualified FPGAs, like the V5QV from Xilinx, are not supported for high-level synthesis. Lastly, we have found that existing HLS tools cannot obtain the same OpenCL code and execute it unchanged on an FPGA, because they need extra code for interfacing between the host and accelerator sides. Moreover, the kernel code needs to undergo major modifications and annotations, so that the generated hardware is efficient, to a degree that {\rremove{won't}} {\radd{will not}} have anything in common with the original OpenCL code. Because of this, the conclusion is that further research on this path is required, but this is beyond the scope of our study, which focuses solely on embedded GPUs, and is more amenable to a future project related to the design of ASIC hardware for the space.

%might be low, as shown by recent studies. 
%Therefore, this is something that needs to be taken into account. 
%Moreover, existing space qualified FPGAs such as Xilinx's V5QV do not support high-level synthesis. Finally, our analysis indicates that the existing HLS tools are unable to get the same OpenCL code and run it unmodified on an FPGA, either because they require additional glue code between the host and accelerator side or because the kernel code has to be heavily modified and annotated, so that the generated hardware is efficient, in a degree that it will have nothing in common with the original OpenCL code. For this reason, we have concluded that investigating further this path is beyond the scope our project, which is focused only on embedded GPUs, and it is more subject to a future project related to ASIC hardware design for space.
%, in addition to the selection of
%products with rad-hard implementation and appropriate performance and SWaP
%(Size, Weight and Power) characteristics for the space domain.

\subsection{GPU Survey Summary}
%\vspace{-0.2cm}

From the four classes of embedded GPUs that we have defined in our taxonomy, we have determined that the FPGA path should no longer be explored during the project to implement commercial COTS GPU designs or open source GPU designs. HLS has potential, but it was considered outside the scope of this project.

%Based on the 4 categories of the embedded GPUs which we have identified in our taxonomy, our analysis indicates that the FPGA path should not be pursuited further for implementing commercial COTS GPU designs, nor open source GPU designs. HLS has potential, but it has been deemed out of the scope of this project.

%Therefore only commercial embedded GPU designs are considered for the next
%project phases, either in their COTS SoC implementations or in radiation hardened ASICs.

Hence, both high-end and low-end products can be used, offering different trade-offs in terms of throughput, power consumption, programming interfaces, complexity, debugging and development tools, open design and functional safety. Based on the information we have collected from vendors, we have chosen a set of on-board GPUs to assess and compare with existing embedded devices as described in the following Section.

%Therefore only commercial embedded GPU devices are considered for the next project phases, either in their COTS SoC implementations or in IP for potential use in future rad-hard ASICs. Both high-end and low-end products can be used, although they offer different trade-offs in performance, energy efficiency, programming interfaces, maturity, development and debugging tools, open specification and functional safety. Based on the additional vendor information we will collect in the near future and the importance of each of the trafeoffs, we will select the most appropriate embedded GPU candidates for evaluation and comparison with existing on-board devices.

\section{Preliminary GPU Evaluation Results} 
\label{sec:results}

\begin{figure}[h]
\centering
\includegraphics[width=\columnwidth]{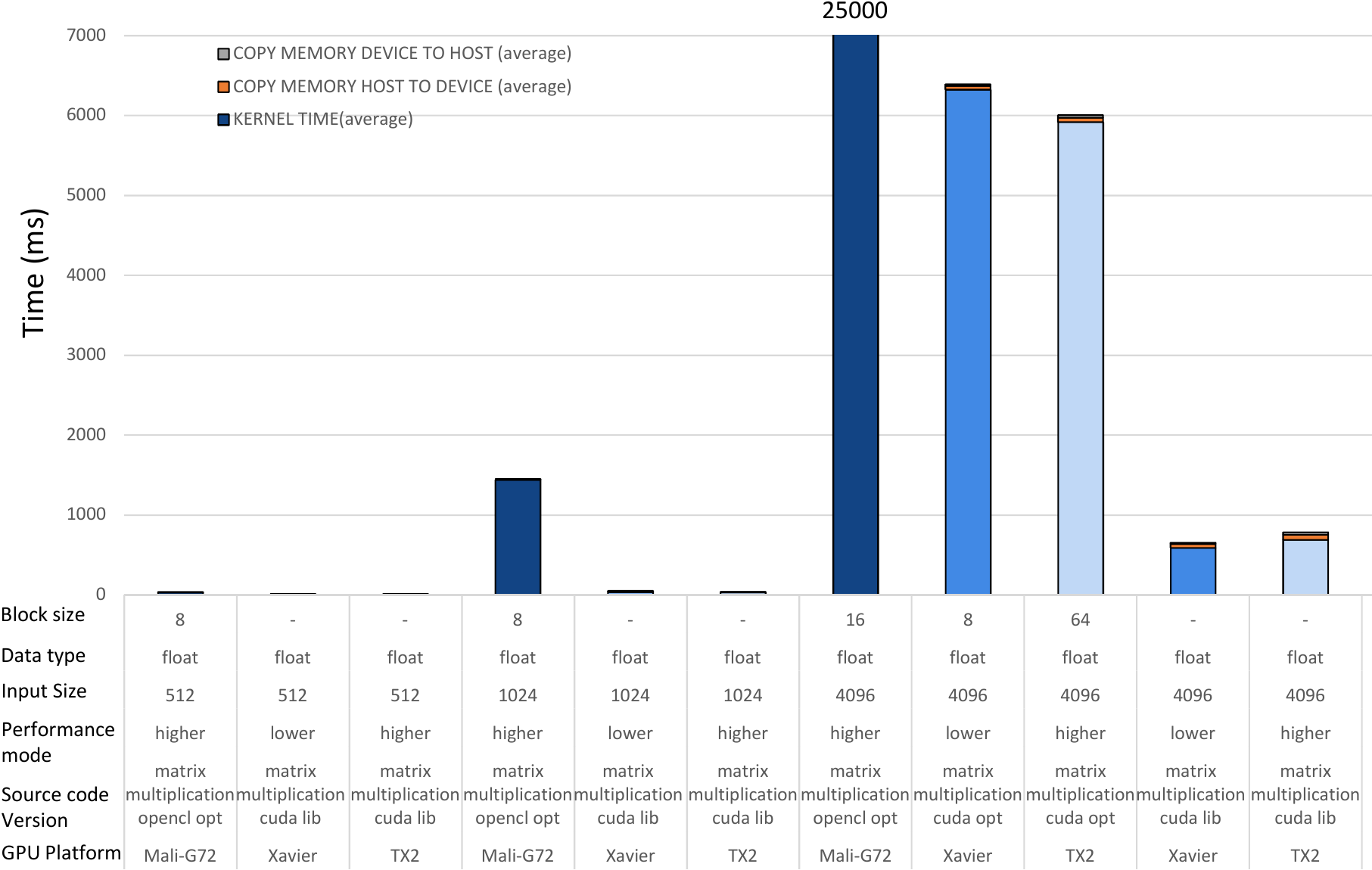}
\caption{\label{fig:sgemm}Performance comparison of various configurations of Matrix Multiplication in the 10W TDP performance mode of the 3 GPU platforms.}
\end{figure}

Based on the above presented surveys on space applications and existing embedded GPUs, we have narrowed down the GPUs to be evaluated.
For cost and effort reasons the selection of the platforms to be evaluated experimentally had to be limited to 3 COTS GPUs, with additional GPU platforms to be evaluated using their published results.
The initial selection included 3 embedded GPUs, two European which can be used either as IP or COTS and the latest non-European GPU which has the highest theoretical performance among all GPUs in the embedded market. In particular, the selected GPUs are:
 the latest ARM Mali GPU G72 found in HiSilicon's HiKey 970, the Imagination Technologies PowerVR Series 6 (GX6650) GPU found in the Renesas RCAR H3 automotive board compatible with ASIL-B functional safety level and the latest embedded NVIDIA platform Xavier which is designed for autonomous driving and certifiable up to the highest automotive safety level, ASIL-D.
All platforms have a 10W TDP performance mode which has been determined as an upper limit for on-board systems based on our analysis and two of them, the RCAR H3 and NVIDIA Xavier are automotive-grade products, designed to comply with functional safety certification{, \radd{therefore they include hardware reliability features}}.

Unfortunately the RCAR H3 production has been canceled by Renesas after its selection, which has not permitted the procurement of this device for benchmarking. For this reason, we have replaced it by another COTS GPU from NVIDIA, the Jetson TX2, which has a product variant specifically designed for industrial applications.

Based on the space application survey we have designed a GPU benchmark suite for space, which consists of kernels extracted from several space domains. Initially we planned to use Brook Auto~\cite{brook_auto} which can facilitate the certification/qualification of the GPU software and allows the programmability of low-end GPUs, however since the selection of the GPU platforms included only the latest high-end GPUs we have used their corresponding native programming APIs, OpenCL and CUDA. More information about the designed benchmark suite is going to be provided in a future publication which is currently under peer review, and the source code will be released as open source after the end of the project. 

\begin{figure}[h]
\centering
\includegraphics[width=\columnwidth]{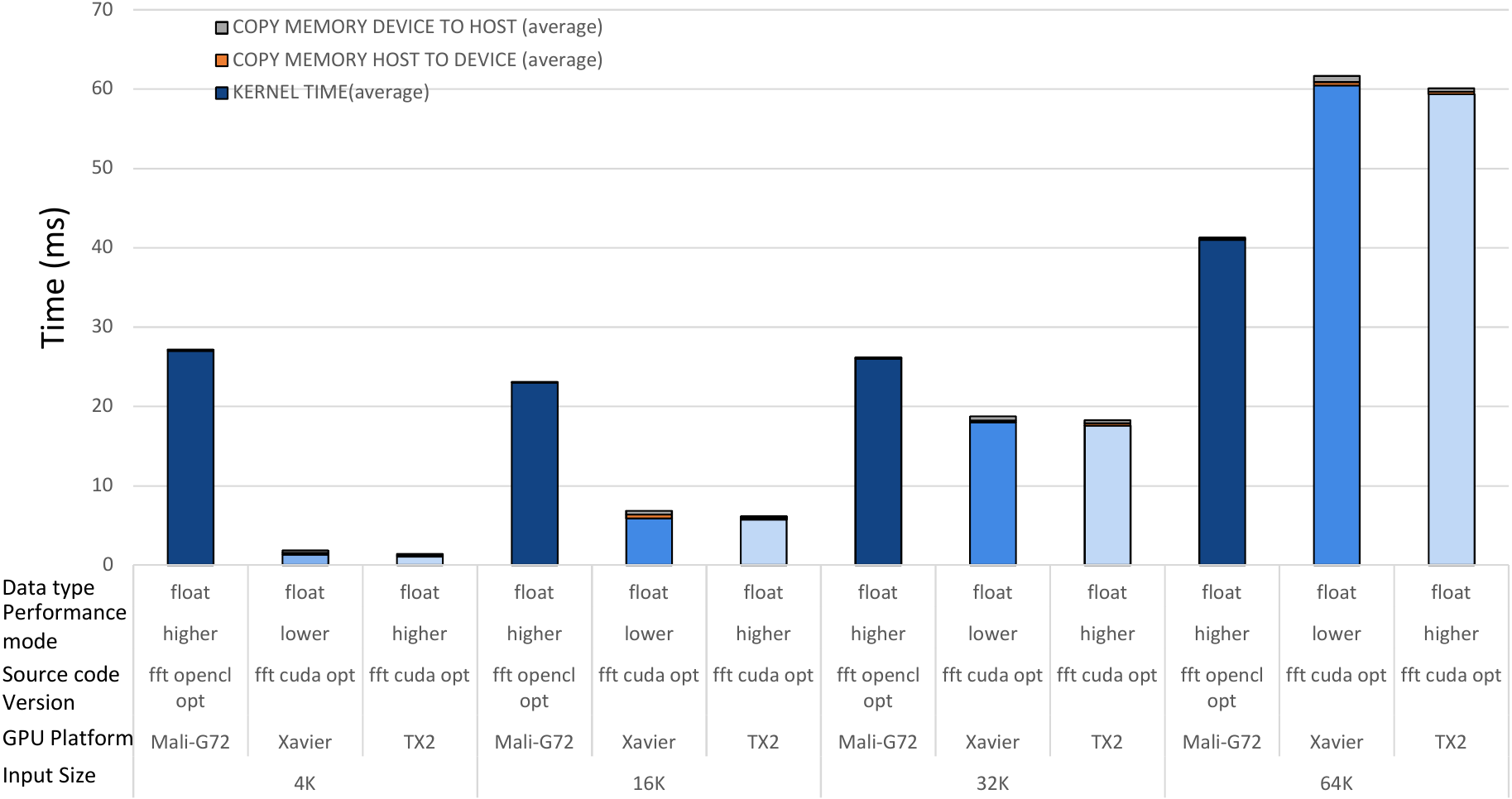}
\caption{\label{fig:fft}Performance comparison of various configurations of Fast Fourier Transform (FFT) in the 10W TDP performance mode of the 3 GPU platforms.}
\end{figure}

In Figure~\ref{fig:sgemm} and Figure~\ref{fig:fft} we present some preliminary results for various implementations of two widely used kernels in space algorithms, the matrix-matrix multiplication and the Fast Fourier Transform (FFT).

In the matrix multiplication benchmark (Figure~\ref{fig:sgemm}) we notice that for small sizes the 3 GPUs have similar performance, however when the size of the input is increased, the NVIDIA GPUs outperform the ARM one. Although this is partially due to the fact that there is no ARM optimised library for this operation, a comparison of the same hand-written and optimised implementation in all platforms shows the same trend. Moreover, an interesting observation is that for the handwritten implementation the TX2 is faster than the Xavier, while when the NVIDIA library is used (CUBLAS), the Xavier is faster.

%Another interesting observation has been that our handwritten and optimised implementation of matrix multiplication for double precision floating point was faster than the NVIDIA's optimised library, CUBLAS on Nvidia Xavier. Although this was unexpected, there are several reasons for that. First, the hardware of this device is not optimised for double precision arithmetic, nor it is its primary workload. As a consequence the NVIDIA library might not be currently optimised for this target. However, in the future this might change, so that in a software update the CUBLAS implementation will probably become faster than our handwritten implementation. Moreover, the design of 

On the other hand, for the FFT (Figure~\ref{fig:fft}) we observe that again for small input sizes the Mali is slower than the NVIDIA GPUs. However, for the large input the hand-written and optimised version performs significantly better in the ARM than in the NVIDIA GPUs.

In short, the preliminary comparison results of the 3 candidate GPU platforms under the same power budget of 10W indicate that there is no embedded GPU platform among them that is definitely better than the others. On the contrary, the results vary based on the actual input sizes, the algorithm and its implementation. For this reason, the entire experimental evaluation has to be completed and all the results to be analysed before reaching a conclusion for the most appropriate GPU platform.

\section{Current and Future Project Work} 
\label{sec:future_work}

{\radd{
\subsection{Further GPU Benchmarking}
}}

Recent results on radiation testing of several embedded GPUs~\cite{troxel} show that embedded GPUs from AMD {\rremove{, which have not been considered in the initial phases of the project due the fact that it is not european technology and at that project stage its newest embedded GPUs with raw performance capabilities in par to Xavier were not available in the market,}} have good reliability properties. {\radd{However, AMD embedded GPUs had not been considered in the initial phases of the project for two reasons. The main reason for this decision was that AMD GPUs do not consitute european technology, which was a primary focus of the project. On the other hand, NVIDIA's GPUs were also not european technology, but they were included as the highest performance embedded GPUs available in the market. Second, in the early project stage when the candidate GPUs were identified, the newest embedded GPUs from AMD which could provide equivalent performance capabilities to the Xavier were not available in the market.}} {\rremove{For this reason}}{\radd{Consequently}}, the detailed comparison among the platforms has been extended in order to include the evaluation of an AMD Embedded Ryzen GPU. This task is currently on-going and it is expected to be completed in some months.

In addition to the evaluation with kernels which we extracted from space applications, we have ported an ESA provided application to CUDA, as a demonstrator that space applications can be parallelised to the GPU programming model. The algorithm we have ported is the Euclid NIR (Near Infrared) image processing algorithm used in ESA's Euclid NIR mission~\cite{euclid_mission}. Our preliminary experimental results on the NVIDIA Xavier platform have been published in~\cite{euclid_nir_gpu} and show that the GPU implementation of the algorithm is 3.5 times faster than the sequential CPU version on the Xavier platform for a standard image size. However for higher resolutions the speedup is 15 times. A comparison with {\rremove{a existing}}{\radd{two representative}} processors used in space, such as LEON 2 and the PowerPC 750 shows an improvement of 806 times and 128 times respectively. Currently we are working on porting the application to OpenCL, so that we can provide a better comparison among the platforms we have considered in the project.

{\radd{
\subsection{Reliability Considerations and Space Adoption Roadmap}
}}

In addition to the performance evaluation and benchmarking, we have started working towards the last main task, which is defining the next steps for the adoption of GPUs in space. In this task, we are mainly focusing on {\rremove{the}} {\radd{performing some preliminary analysis of}} reliability solutions {\rremove{of}} {\radd{for}} embedded GPUs. {\radd{Recall that GPU4S is the first ESA project on embedded GPUs, so its purpose is not to address all problems related to embedded GPUs in space. Instead it rather aims to assess in a first stage whether they can be beneficial for the space domain and identify potential issues that need to addressed for their adoption, providing some guidelines in this direction.}} 

As already mentioned, approaches for other domains such as the automotive are usually employed by the space domain in order to reduce non-recurrent costs. Regarding COTS GPUs we propose the use of a solution that we previously developed for the automotive sector, based on dual-modular software-based diverse redundancy~\cite{sergi_iolts}. This solution exploits the parallelism of the GPU hardware in order to execute two copies of the same GPU kernel, while ensuring that they are executed on different hardware resources and with a time slack between them, so that a soft error cannot be manifested in both copies in the same way. The comparison of the results of the two copies is performed by a safety microcontroller, such the one already included in automotive-grade platforms like Xavier.

For the GPU IP solutions, since their hardware design can be modified, we propose a hardware solution, again from our previous work in the automotive domain~\cite{sergi_date}. This solution applies the same concept with~\cite{sergi_iolts}, however in this case it is the hardware which guarantees the spatial and temporal diversity. In particular, we present two hardware scheduler modifications. In the first policy we partition the GPU so that each kernel version do not use the same resources, while in the second we follow a round-robin dispatch of threads of the kernel, but again we guarantee that the two copies are not executed in the same hardware elements.

{\radd{It is worth noting that most IP GPUs are available in RTL form and are not tied to any particular technology or library implementation, although some companies such as ARM provide also cell libraries optimised for different purposes, like high-performance, low-power or area.
This means, that these GPU IPs can be synthesized with appropriate cell libraries for space such as radiation hardened or the naturally immune to latch-up ST Microelectronics' 28nm FDSOI european technology, currently used in the ARM-based DAHLIA H2020 project which targets the space domain. However, it is not possible to make any trustworthy estimations about how GPU designs will perform in such techonology nodes or older space technologies such as 65nm. We expect that in that case, the obtained frequency of the embedded GPUs with older technologies will be slower, but this might not be an issue since space electronics already need to be operating in lower frequencies in order to keep power and thermal dissipation within the on-board specification limits. Moreover, the area is inevitably expected to be larger and the energy efficiency to be lower than the ones obtained for free with the latest technology nodes, thanks to Moore's law and Dennard scaling. However, we should not forget that embedded GPUs have been in the market for long time now, as we have mentioned in Introduction, so there is enough evidence of embedded GPUs implemented in such older manufacturing processes and still provide benefits over their CPU counterparts in the same SoCs.
}}

{\radd{
\section{Challenges and Difficulties}
\label{sec:challenges}

In this section, we describe the challenges and the difficulties we have faced in the project so far. The most important challenge we had to face was the limited information provided by the GPU IP companies. In particular, semiconductor industries are very protective about the details of their designs. The situation is even more complicated with embedded GPU designs, whose internals are very rarely provided, even in GPUs with available open source drivers such as Broadcom's VideoCore IV and AMD GPUs. For this reason, we had to perform our initial hardware survey considering only the publicly available information found in marketing material. The next step was to establish communication channels with all the GPU vendors, explain the purpose of our analysis and sign several non-disclosure agreements (NDAs) with each of them, in order to obtain non-public information about their products. However, this is a very long administrative process which requires the involvement of the legal departments of large companies and institutions and has caused long, unexpected delays in the project. In addition, although we obtained such exclusive information which helped us further for choosing the most promising embedded GPU candidates for benchmarking, not all rationale behind this selection can be disclosed in documents such as this publication.

Another limitation when dealing with semiconductor industries, is that no price information or internal data related to performance, power and area is provided, until the end user signs a customer agreement and specifies exactly the requirements of the design eg. which IP would like to license, its configuration parameters etc. However, this was not possible in our case and prevented us to make a comprehensive selection of GPU IP including a comparison of these properties for different available IPs and to be able to perform an accurate normalisation of projected power, performance and area of these designs on older space technologies.

Another important difficulty we faced was the unavailability of some candidate embedded GPU platforms as we already commented. In particular, the initially selected RCAR H3 from Renesas was out of stock at the moment we placed a purchase order. After one year of waiting time, the manufacturer decided to cease its production and cancel all pending orders. This change did not allow Imagination Techonologies' GPU, one of two european IPs which was selected for evaluation to be represented in our benchmarking, caused delays in our evaluation phase and forced us to look for an alternative platform to keep a minimum of 3 boards for experimental evaluation. Similarly, the unavailability of platforms based on the latest embedded AMD GPUs at the time of platform selection kept initially AMD products out of consideration.

Finally, another challenge we had to cope with was the budget limitation, for a project that we extended well beyond its narrow initial scope. Planned as a pilot project with an initial budget of 150K euros, the number of possible GPUs to be purchased and experimentally evaluated was limited to 3. Moreover, this further constrained the effort that could be devoted to conduct the various surveys including communication and legal agreements with vendors, platform setting up, code development and optimisation in order to perform a fair comparison and a demonstrator, as well the preliminary reliability assessment and definition of the adoption roadmap. Despite this limitation, we managed to extend the project outcome significantly beyond its limited scope, for example covering much more hardware devices that the four classes of embedded GPUs we identified in our taxonomy: high-level synthesis with OpenCL on FPGAs and machine learning accelerators. Moreover we proposed additional GPU-oriented reliability solutions inspired from the use of embedded GPUs in other safety critical systems, instead of relying only on legacy solutions used for general purpose COTS hardware in space.
}}

%\vspace{-0.1cm}
\section{Summary}
\label{sec:concl}

In this paper we provide an overview of the objectives and the results we have obtained so far in the GPU4S project, which explores the suitability of embedded GPUs in the space domain.

%In this paper we described the goals and the preliminary results of the GPU4S project, which studies the applicability of embedded GPUs in the space domain.

Based on the results of our first survey of space applications and domains, we have since identified that embedded GPUs are suitable for a variety of multi-domain algorithms such as vision-based navigation, image processing, neural network processing, and signal processing, and that, depending on the specific mission, several characteristics such as reliability and thermal requirements are relevant to the selection of candidate GPUs.

%From our early survey results in space applications and domains, we have identified so far that GPUs are appropriate for a wide range of algorithms from several domains like  vision based navigation, image processing, neural network processing and signal processing, and that depending on the particular mission, there are different characteristics like reliability requirement and thermal which need to be taken into account for the selection of the candidate GPUs.

In terms of the domain of embedded GPUs, there are several IP options of the embedded GPU domain that are potential candidates, most of them European, but each of them has different trade-offs that must be evaluated in the next project stages, in order to proceed with the final platform choice that will be experimentally evaluated. However, our research suggests that soft GPU IP solutions in FPGA, as well as high-level synthesis, are not suitable for further investigation in this project (HLS) or at all (soft GPUs).

%Regarding the embedded GPU domain we have seen that several embedded GPU IP options are potential candidates, most of which European, but each one offers different trade-offs which have to be further evaluated in the next steps of the project, once all information is obtained, in order to perform the final selection of the platforms that will be evaluated experimentally. However, our analysis indicates that soft GPU IP solutions on FPGA as well as high-level synthesis are not appropriate for further exploration in this project (HLS) or at all (soft GPUs).
%\vspace{-0.1cm}

Our preliminary experimental results under the same 10W power budget indicate that there is not a clear performance advantage of one embedded GPU over the others in our candidate list. In particular, the results depend on the input size of the algorithm, the algorithm and its implementation, so a complete analysis is required with the rest of the benchmarks we have developed in our benchmark suite. The results of this task are expected to be included in a future publication which is currently under peer review, together with detailed information about our benchmark suite, which is going to be released under an open source license. However preliminary results with a space case study ported to an embedded GPU and compared with existing space processors, show clearly that space algorithms can be a good fit for GPUs and that embedded GPUs can provide the desired required performance.

Finally on the reliability side we have explored the use of proposals from the automotive domain both at software level, in case COTS GPUs are used, as well as at hardware level when a GPU IP solution is adopted.

%\selectfont

\section*{Acknowledgments}
%The research leading to these results has received funding from the European
%Space Agency under the GPU4S Project, XXX.
%This work has also been partially supported by the Spanish Ministry of Science and Innovation under grant TIN2012-34557 and the HiPEAC Network of Excellence. 
%Leonidas Kosmidis has been partially supported by XXX and Jaume Abella has been partially supported by the Ministry of Economy and Competitiveness under Ramon y Cajal postdoctoral fellowship number RYC-2013-14717.
This work has received funding from the the European
Space Agency (ESA) under the GPU4S (GPU for Space) Project, answer to the ESA ITT AO/1-9010/17/NL/AF tender with title
"Low Power GPU Solutions For High Performance On-Board Data Processing" and from the European Research Council (ERC) under the European Union's Horizon 2020 research and innovation programme (grant agreement No. 772773). This work
has also been partially supported by the Spanish Ministry of Economy and Competitiveness (MINECO) under grant TIN2015-65316-P
and the HiPEAC Network of Excellence. MINECO partially supported Leonidas Kosmidis under Juan de la Cierva Formaci\'{o}n postdoctoral fellowship (FJCI-2017-34095) and Jaume Abella under Ramon y Cajal postdoctoral fellowship (RYC-2013-14717).

%\linespread{1}
%\small
%\selectfont
\bibliographystyle{IEEEtran}
\bibliography{IEEEabrv,biblio}

\end{document}